\documentclass[sigconf, natbib=true]{acmart}
\usepackage{subcaption}
\usepackage{multirow}
\AtBeginDocument{%
  }

\renewcommand\footnotetextcopyrightpermission[1]{} % removes footnote with conference information in first column
\acmConference[]{}{}{} 
\acmISBN{}
\acmDOI{}
\begin{document}

%%
%% The "title" command has an optional parameter,
%% allowing the author to define a "short title" to be used in page headers.
\title{Aligning Dense Retrievers with LLM Utility via Distillation}
% Utility Aligned Embeddings: 
%%
%% The "author" command and its associated commands are used to define
%% the authors and their affiliations.
%% Of note is the shared affiliation of the first two authors, and the
%% "authornote" and "authornotemark" commands
%% used to denote shared contribution to the research.
\author{Rajinder Sandhu}
\email{rajinder@layer6.ai}
\affiliation{%
  \institution{Layer 6 AI}
  \city{Toronto}
  \state{ON}
  \country{Canada}
}
\author{Di Mu}
\email{di@layer6.ai}
\affiliation{%
  \institution{Layer 6 AI}
  \city{Toronto}
  \state{ON}
  \country{Canada}
}
\author{Cheng Chang}
\email{jason@layer6.ai}
\affiliation{%
  \institution{Layer 6 AI}
  \city{Toronto}
  \state{ON}
  \country{Canada}
}
\author{Md Shahriar Tasjid}
\email{tasjid@dal.ca}
% \authornotemark[1]
\affiliation{%
  \institution{Dalhousie University}
  \city{Halifax}
  \state{NS}
  \country{Canada}
}
\author{Himanshu Rai}
\email{himanshu@layer6.ai}
\affiliation{
  \institution{Layer 6 AI}
  \city{Toronto}
  \state{ON}
  \country{Canada}
}
\author{Maksims Volkovs}
\email{maks@layer6.ai}
\affiliation{
  \institution{Layer 6 AI}
  \city{Toronto}
  \state{ON}
  \country{Canada}
}
\author{Ga Wu}
\email{ga.wu@dal.ca}
% \authornotemark[1]
\affiliation{%
  \institution{Dalhousie University}
  \city{Halifax}
  \state{NS}
  \country{Canada}
}

%%
%% By default, the full list of authors will be used in the page
%% headers. Often, this list is too long, and will overlap
%% other information printed in the page headers. This command allows
%% the author to define a more concise list
%% of authors' names for this purpose.
% \renewcommand{\shortauthors}{Trovato et al.}

%%
%% The abstract is a short summary of the work to be presented in the
%% article.
\begin{abstract}
  
Dense vector retrieval is the practical backbone of Retrieval- Augmented Generation (RAG), but similarity search can suffer from precision limitations.
Conversely, utility-based approaches leveraging LLM re-ranking often achieve superior performance but are computationally prohibitive and prone to noise inherent in perplexity estimation. We propose Utility-Aligned Embeddings (UAE), a framework designed to merge these advantages into a practical, high-performance retrieval method. We formulate retrieval as a distribution matching problem, training a bi-encoder to imitate a utility distribution derived from perplexity reduction using a Utility-Modulated InfoNCE objective. This approach injects graded utility signals directly into the embedding space without requiring test-time LLM inference. On the QASPER benchmark, UAE improves retrieval Recall@1 by 30.59\%, MAP by 30.16\% and Token F1 by 17.3\% over the strong semantic baseline BGE-Base. Crucially, UAE is over 180$\times$ faster than the efficient LLM re-ranking methods preserving competitive performance, demonstrating that aligning retrieval with generative utility yields reliable contexts at scale.
\end{abstract}


% \received{20 February 2007}
% \received[revised]{12 March 2009}
% \received[accepted]{5 June 2009}

%%
%% This command processes the author and affiliation and title
%% information and builds the first part of the formatted document.
\maketitle

\section{Introduction}
\label{sec:intro}

Dense vector similarity search remains the bedrock of real-world Retrieval-Augmented Generation (RAG) systems. By mapping queries and candidates into a shared representation space, these systems leverage efficient Approximate Nearest Neighbor (ANN) search to handle large-scale datasets with minimal latency~\cite{es23,florin24,oz24}. However, this paradigm is increasingly criticized for its reliance on semantic similarity as a proxy for generative utility. Growing evidence suggests that passages with high semantic similarity (topical overlap) often fail to provide answer-critical information and can even introduce semantic distractors that mislead the generator, especially in long-context settings where incorrect but similar passages increase decoding uncertainty~\cite{florin24,ori23,bowen24,fangzheng25}.

\begin{figure}[t]

    \centering
    \includegraphics[width=0.8\columnwidth]{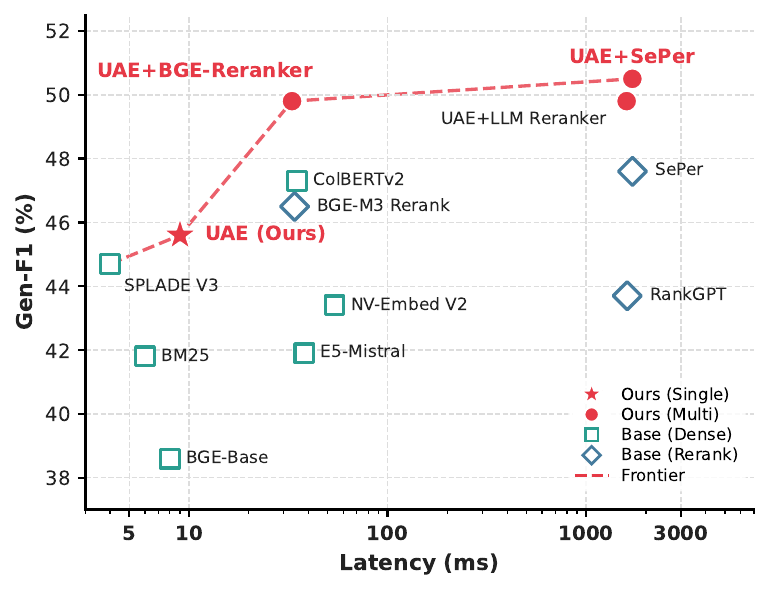}
    \vspace{-3mm}
    \caption{\textbf{Efficiency vs. Performance.} UAE (Read Star) occupies the optimal sweet spot: it approaches the performance of strong baselines while being $\sim$100x faster.}
    \label{fig:efficiency}
\vspace{-2mm}
\end{figure}

To bridge this gap, current state-of-the-art approaches shift toward utility-based retrieval \cite{yi25, dai2025seper, tian2026predicting, UtilityQwen2025}, where relevance is defined by how effectively a document helps a Large Language Model (LLM) produce a correct response~\cite{oz24}. In practice, this is often measured via perplexity reduction: a document is considered useful if its presence as context makes the ground-truth answer more predictable to the model \cite{yi25}. While conceptually sound, these utility-based approaches face a significant practicality wall. Relying on LLMs for query generation or post-hoc re-ranking is computationally prohibitive for large-scale deployment~\cite{alireza24,zixuan24,yi25}. Furthermore, utility signals derived from perplexity are notoriously noisy and stochastic, sensitive to token-level variations and decoding dynamics that make them difficult to use as stable training targets~\cite{tian2026predicting, holtzman2020curious}. This necessitates complex, multi-stage architectures that improve performance at the cost of extreme inference latency and high computational overhead~\cite{yucheng24,hao25}.

\begin{figure*}[t]
    \centering
    \includegraphics[width=0.95\linewidth]{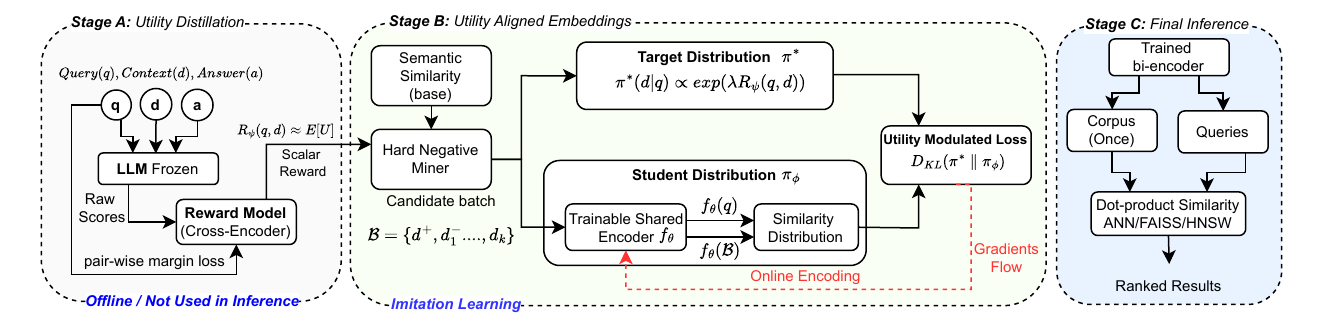}
    \vspace{-3mm}
    \caption{
    Overview of Utility-Aligned Embeddings (UAE).
    Utility is distilled offline into a reward model (Stage A), which defines a target utility distribution used to align a dense retriever via distribution matching (Stage B). At inference time, the trained bi-encoder supports standard ANN retrieval without any reward model or LLM inference (Stage C).
    }
    \label{fig:overview}
\end{figure*}

In this work, we propose Utility-Aligned Embeddings (UAE), a framework designed to merge the efficiency of dense retrieval with the superior performance of utility-based methods. Our core insight is that generative utility should be distilled directly into the bi-encoder’s (dual-encoder dense retriever) embedding space, bypassing expensive test-time LLM inference while capturing the generator’s actual preferences. We formulate this alignment as a distribution matching problem; rather than treating noisy utility scores as direct regression targets (which can lead to unstable learning and overfitting~\cite{ghosh2017robust, shao2021sequence}) we train the bi-encoder to imitate a utility-induced target distribution via a Utility-Modulated InfoNCE objective (Figure~\ref{fig:overview}). Our empirical results on the QASPER benchmark~\cite{dasigi2021dataset} demonstrate that UAE significantly outperforms standard semantic retrievers \cite{xiao2023cpack}, improving reterival Recall@1 by 30.59\% and Token F1 by 17.3\%. Crucially, UAE occupies a unique Pareto-optimal position: it provides substantial performance gains while being 180$\times$ faster than the efficient LLM-based re-ranking methods \cite{sun-etal-2023-chatgpt}. Figure~\ref{fig:efficiency} highlights the advantage of the proposed method. By maintaining standard ANN compatibility, UAE offers a practical, scalable solution for building high-fidelity RAG systems that are both informative and computationally efficient.

\begin{figure}[t]
     \centering
     \includegraphics[width=0.93\columnwidth]{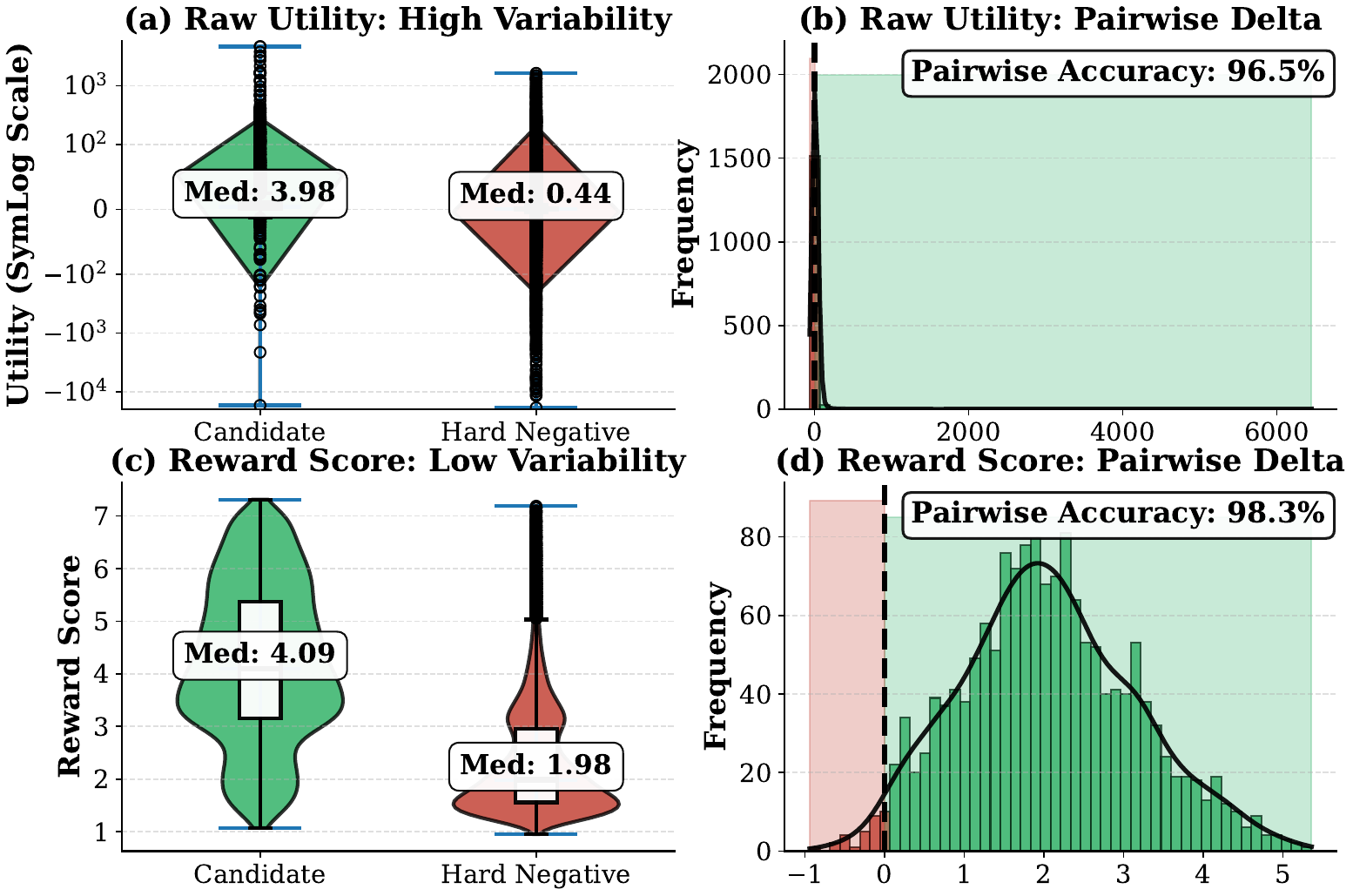}
     \caption{Raw generative utility scores (top) are heavy-tailed and highly variable, making direct regression unstable and motivating reward-based smoothing. Reward model learned (bottom) mitigates the problem.}
     \Description{ Distributional instability of raw generative utility.}
\label{fig:utility_analysis}
\end{figure}

\section{Utility Aligned Embeddings}
\label{sec:method}

Our objective is to develop a dense retrieval system where the embedding space structure is supervised by generative utility rather than simple semantic overlap. However, the high variance and heavy-tailed nature of raw utility scores (Figure \ref{fig:utility_analysis} top) make them unsuitable as direct supervision targets for a bi-encoder. To bridge this gap, we decompose the alignment process into a two-stage distillation framework (Figure \ref{fig:overview}). First, we stabilize the noisy, computationally expensive utility signals by distilling them into a parameterized reward model. Second, we align the dense retriever with this model by training it to imitate the resulting utility-induced ranking distribution through supervised distribution matching. 

\subsection{Parameterized Utility Approximation}
\label{subsec:reward_approx}
The utility of a context document $d$ for correctly answering a question $q$ is denoted as $U(q, d, a)$, where $a$ represents the ground-truth answer. In practice, utility is typically estimated via perplexity:
\begin{equation}
U(q, d, a) = \exp \left( \frac{1}{|a|} \sum_{t=1}^{|a|} \log p_\theta(x_t \mid x_{<t}, q, d) \right),
\label{eq:utility}
\end{equation}
where $p_\theta$ represents the token-level generation probability of a language model. Given that natural language allows for multiple valid expressions of the same answer $a \in \mathcal{A}_q$, we extend this definition to the expected utility $U(q, d, \mathcal{A}_q) = \mathbb{E}_{a \in \mathcal{A}_q} [U(q, d, a)]$. Theoretically, this utility function is deterministic relative to the finite set of ground-truth answers $\mathcal{A}_q$. In practice, however, the estimation is highly stochastic and fragile; The instability arises from three primary factors: (1) the intractable variety of linguistic expressions for ground-truth answers; (2) non-deterministic model outputs caused by matrix parallelism and floating-point associativity errors; and (3) a high sensitivity to answer length $|a|$. 

Directly using raw generative utility through regression (using standard objectives like Mean Squared Error or Huber losses) is unreliable due to the aforementioned stochasticity and heavy-tailed distribution. Instead, we formulate reward learning as a pairwise ranking problem, focusing exclusively on preserving the generator’s ordinal preferences. We construct training quadruplets $(q, \mathcal{A}_q, d_i, d_j)$ and optimize a parameterized reward model $R_\psi$ using a margin-based ranking loss:
\begin{equation}
    \mathcal{L}_{\text{Reward}} = \max(0, \delta - (R_\psi(q, d_i) - R_\psi(q, d_j))),
\end{equation}
where we enforce $R_\psi(q, d_i) > R_\psi(q, d_j) + \delta$ whenever $U(q, d_i, \mathcal{A}_q) > U(q, d_j, \mathcal{A}_q)$. By prioritizing the relative ordering of candidates over the approximation of exact utility values, this formulation provides a more robust signal for learning utility-driven representations. Here, $R_\psi$ is a Transformer-based encoding model (BERT). As illustrated in Figure~\ref{fig:utility_analysis} (bottom), this learned reward model effectively discriminates between truly useful contexts and the "hard semantic negatives" that often deceive standard top-1 similarity search.

\subsection{Reward-Guided Embeddings Training}
\label{subsec:retriever_training}
The reward model $R_\psi$ effectively captures utility but, as a cross-encoder, is computationally expensive for large-scale inference. We therefore use $R_\psi$ as an offline teacher to distill utility preferences into a dense bi-encoder, preserving ANN search efficiency. Rather than using reinforcement learning (which is destabilized by massive discrete action spaces and high reward variance), we adopt an imitation learning formulation. We treat the normalized reward distribution as a target expert policy $\pi^*$ and train the retriever $\pi_\phi$ to minimize their divergence via supervised gradients, bypassing the exploration challenges of traditional RL.

\noindent{\bf Distribution Matching Objective.}
To align the retriever with generative utility, we define a target (teacher) distribution $\pi^*$ and a retriever (student) distribution $\pi_\phi$ over a set of candidate contexts $\mathcal{B}$ as follows:
\begin{align}
    \pi^*(d_i \mid q) &= \frac{\exp(\lambda \cdot R_\psi(q, d_i))}{\sum_{d_j \in \mathcal{B}} \exp(\lambda \cdot R_\psi(q, d_j))}, \\
    \pi_\phi(d_i \mid q) &= \frac{\exp(\langle f_\phi(q), f_\phi(d_i) \rangle / \tau)}{\sum_{d_j \in \mathcal{B}} \exp(\langle f_\phi(q), f_\phi(d_j) \rangle / \tau)},
\end{align}
where $\lambda$ is a temperature hyperparameter controlling the sharpness of the target utility distribution and $\tau$ is the retriever's scaling temperature. The retriever is optimized by minimizing the Kullback-Leibler (KL) divergence between the two distributions:
\begin{equation}
\begin{aligned}
    \mathcal{L}_{\text{UAE}}(\phi) &= \mathbb{E}_q \left[ D_{\text{KL}}(\pi^* \parallel \pi_\phi) \right] \\ &\propto - \sum_q \sum_{d_i \in \mathcal{B}} \pi^*(d_i \mid q) \log \pi_\phi(d_i \mid q).
\end{aligned}
\vspace{-1mm}
\end{equation}
This objective reshapes the embedding space to reflect the generator's preferences: high-utility documents are pulled toward the query, while low-utility contexts are pushed away. Critically, unlike standard InfoNCE which treats all negatives as equally irrelevant, this formulation preserves the graded, ordinal structure of the utility space, allowing the retriever to distinguish between varying degrees of relevance.

\noindent{\bf Utility-Aware Hard Negative Mining.}
Since computing the full distribution over the entire corpus is computationally intractable, we adopt the Noise Contrastive Estimation (NCE) paradigm to approximate the global distribution using a combination of gold contexts and informative negative samples. However, random negatives often provide trivial gradients, while standard semantic hard negatives may accidentally include documents that the generator finds useful. To address this, we implement a Utility-Gated Mining, where a document $d^-$ is selected as a negative only if it satisfies two conditions: (1) it is semantically similar to the query (ranked in the top-$k$ by a base similarity function $S_{\mathrm{sem}}$), and (2) it is assigned substantially lower utility than the gold evidence $d^+$ by the reward model, satisfying $R_\psi(q, d^+) - R_\psi(q, d^-) > \delta$. The margin $\delta$ governs the trade-off between coverage and signal purity; a conservative margin ensures that the retriever focuses on resolving \textit{semantic distractors}, documents that are proximally located in embedding space but offer negligible generative utility. This gated approach, combined with the robustness of the KL-divergence objective, prevents the propagation of noise from the reward model to the final embeddings.

\noindent{\bf Scalability.} 
To ensure training efficiency, we adopt an in-batch encoding protocol similar to DPR \citep{karpukhin-etal-2020-dense}. Rather than re-encoding the entire corpus, we dynamically encode only the sampled candidate set $\mathcal{B} = \{d^+, d^-_1, \dots, d^-_k\}$ for each query batch. This enables end-to-end optimization of the embedding space with gradients flowing through a shared encoder, while keeping training costs independent of the total corpus size. Crucially, the expensive reward model is utilized only during this offline phase; at inference time, the corpus is indexed for standard ANN search, fully decoupling utility alignment from runtime latency.

\section{Experiments}
\label{sec:experiments}
\noindent\textbf{Datasets \& Evaluation Protocol.} We evaluate UAE on two distinct RAG benchmarks: QASPER~\cite{dasigi2021dataset} (long-doc scientific QA) and NewsQA~\cite{trischler2017newsqa} (short-doc news extraction). 
We employ a hard-negative setting where the candidate pool ($N=50$) for each query is constructed via dense retrieval (BGE-Base~\cite{xiao2023cpack}) and reward model utility. This populates the pool with the semantic distractors described in Section \ref{subsec:retriever_training}, rigorously testing the model's ability to prioritize true generative utility over high-similarity non-answers.

\noindent \textit{Generation Protocol.}
To measure downstream utility, we utilize Llama-3-8B-Instruct~\cite{dubey2024llama} as the fixed generator with greedy decoding (temperature=0) for reproducibility. Dataset-specific system prompts align the generator’s output with the ground-truth format: extractive phrases for NewsQA and evidence-based summaries for QASPER. Performance is quantified using Token F1 \cite{rajpurkar-etal-2016-squad} and ROUGE-L~\cite{lin2004rouge} to assess both informational accuracy and structural fluency.

\noindent \textit{Model Configuration.} We initialize the utility reward model with microsoft/deberta-v3-base \cite{he2021deberta} and the retriever with BAAI/bge-base-en-v1.5 \cite{xiao2023cpack}. We also employ Low-Rank Adaptation (LoRA) \cite{hu2022lora} for parameter-efficient fine-tuning across both components.
\vspace{1mm}

\begin{table*}[t]
\centering
\caption{Comparison of UAE against baselines on QASPER and NewsQA. 
We report Recall@1 (\textbf{R@1}), Recall@3(\textbf{R@3}), Expected Utility @1 (\textbf{ExpUtil@1}), Mean Average Precision (\textbf{MAP}), and Generation Metrics (\textbf{Gen-F1}, \textbf{ROUGE-L}). 
Efficiency is measured as latency (Lat.) in milliseconds. N/A denotes not available.}
\label{tab:main_results}
\vspace{-3mm}
\setlength{\tabcolsep}{3.5pt} 
\renewcommand{\arraystretch}{1.1} 
\small 
\resizebox{\textwidth}{!}{%
\begin{tabular}{cl ccccccr ccccccr} 
\toprule
&& \multicolumn{7}{c}{\textbf{QASPER} (Long-doc)} & \multicolumn{7}{c}{\textbf{NewsQA} (Short-doc)} \\
\cmidrule(lr){3-9} \cmidrule(lr){10-16} 
&\textbf{Method} & \textbf{R@1} & \textbf{R@3} & \textbf{ExpUtil@1} & \textbf{MAP} & \textbf{Gen-F1} & \textbf{ROUGE-L} & \textbf{Lat.} & \textbf{R@1} & \textbf{R@3} & \textbf{ExpUtil@1} & \textbf{MAP} & \textbf{Gen-F1} & \textbf{ROUGE-L} & \textbf{Lat.} \\
\midrule
\multirow{5}{*}{\rotatebox[origin=c]{90}{\parbox{1.2cm}{\centering Classic \\ Retrieval}}}
&BM25 & 26.26 & 49.70 & 0.6792 & 27.81 & 20.5 & 20.7 & 5 & 49.61 & 69.79 & 5.1859 & 52.13 & 41.8 & 42.2 & 6 \\
&SPLADE V3 \cite{lassance2024spladev3} & 38.46 & 61.76 & 0.9836 & 32.13 & 25.2 & 25.2 & 5 & 50.58 & 73.67 & 5.3588 & 53.34 & 44.7 & 45.2 & 4 \\
&InfoNCE Tuned & 6.21 & 18.23 & 0.0932 & 8.61 & 5.9 & 6.0 & 9 & 22.45 & 51.29 & 4.1819 & 32.60 & 25.5 & 25.9 & 9 \\
&BGE-Base \cite{xiao2023cpack} & 36.87 & 61.28 & 0.8201 & 38.02 & 23.9 & 24.1 & 8 & 34.15 & 57.37 & 4.7378 & 43.84 & 38.6 & 39.0 & 8 \\
&ColBERTV2 \cite{santhanam-etal-2022-colbertv2} & 40.75 & 64.87 & 0.9790 & 33.98 & 26.7 & 26.8 & 46 & 55.11 & 76.53 & 5.5024 & 58.31 & 47.3 & 47.8 & 35 \\
\midrule
\multirow{2}{*}{\rotatebox{90}{\parbox{1cm}{\centering \small LLM Dense \\ Retriever}}}
&NV-Embed V2 \cite{lee2024nv} & 47.04 & 71.75 & 0.9749 & 38.67 & 27.8 & 27.7 & 52 & 47.35 & 72.06 & 5.3325 & 51.28 & 43.4 & 43.9 & 54 \\ \addlinespace[0.8ex]
&E5-Mistral \cite{UtilityQwen2025} & 40.46 & 65.24 & 0.8688 & 33.93 & 26.0 & 26.1 & 39 & 45.15 & 69.99 & 5.2489 & 47.64 & 41.9 & 42.5 & 38 \\ \addlinespace[0.8ex]
\midrule
\multirow{5}{*}{\rotatebox[origin=c]{90}{\parbox{1.55cm}{\centering Multi-Stage \\ (LLM) Rerank}}}
&BGE-V2-M3 reranker \cite{chen2024bge} & 43.05 & 66.79 & 1.0882 & 42.93 & 27.4 & 27.5 & 52 & 60.67 & 77.62 & 5.7339 & 65.49 & 46.5 & 47.0 & 34 \\
&RankGPT \cite{sun-etal-2023-chatgpt} & 48.89 & 71.52 & 1.2328 & 46.13 & 29.2 & 29.3 & 1663 & 49.68 & 71.54 & 4.9680 & 46.78 & 43.7 & 44.2 & 1610 \\
&GainRAG \cite{yi25} & 46.30 & 69.23 & 1.3057 & 45.29 & 28.0 & 27.9 & 2293 & 41.20 & 64.17 & 5.0329 & 50.96 & 40.7 & 41.4 & 2104 \\
&UtilityQwen \cite{UtilityQwen2025} & 40.95 & 64.25 & 1.096 & 37.66 & 24.0 & 24.1 & 3200 & 42.77 & 65.35 & 5.013 & 47.90 & 38.9 & 39.3 & 3194 \\
&SePer \cite{dai2025seper} & 59.84 & 72.71 & 2.3445 & 50.12 & 32.6 & 32.1 & 1696 & 60.61 & 78.65 & 6.1131 & 66.17 & 47.6 & 48.0 & 1711 \\
\midrule
&\textbf{UAE (Ours)} & 48.15 & 70.56 & 1.3269 & 49.49 & 27.0 & 27.0 & 9 & 54.90 & 77.30 & 5.8184 & 62.31 & 45.6 & 46.1 & 9 \\
\midrule
\midrule
\multirow{3}{*}{\rotatebox[origin=c]{90}{\parbox{1.15cm}{\centering Rerank \\ Ablation}}}
&UAE + BGE-reranker & 44.45 & 70.71 & 1.1458 & 45.94 & 28.5 & 28.5 & 43 & 65.65 & 86.22 & 5.8727 & 72.04 & 49.8 & 50.3 & 33 \\
&UAE + SePer & 61.91 & 74.11 & 2.4194 & 53.39 & 34.0 & 33.3 & 1704 & 66.43 & 88.10 & 6.2394 & 73.15 & 50.5 & 51.0 & 1709 \\
&UAE + LLM Reranker & 50.48 & 75.70 & 1.2631 & 50.65 & 29.8 & 29.7 & 1617 & 60.87 & 87.32 & 5.7416 & 67.92 & 49.8 & 50.3 & 1599 \\
\bottomrule
\end{tabular}}

\end{table*}

\begin{figure}[t]
    \centering
    \vspace{-0.3cm}
    \includegraphics[width=\columnwidth]{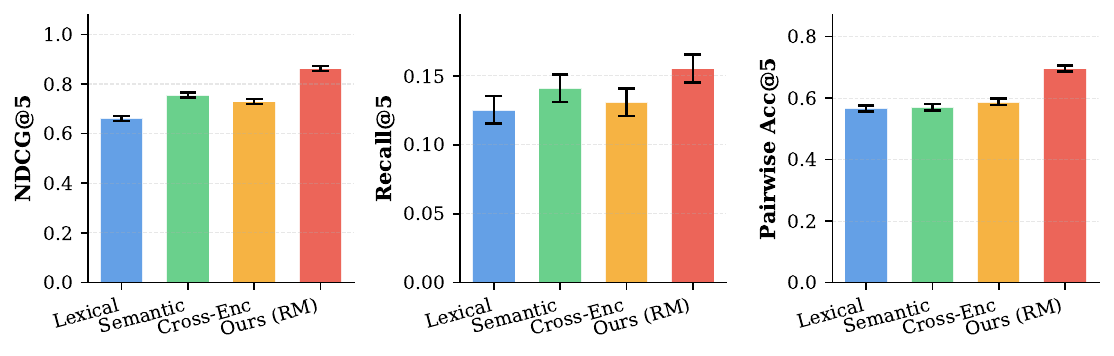}
    \vspace{-5mm}
    \caption{Alignment of various retrieval models with the LLM's utility distribution. Reward Model (RM), while not a practical retriever, significantly outperforms lexical, semantic, and cross-encoder baselines across all ranking metrics, providing a higher-fidelity supervision signal for distillation.}
    \label{fig:signal_analysis}
    \vspace{-3mm}
\end{figure}

\vspace{1mm}
\noindent{\bf Reward Model Validation.} We verify the distillation fidelity of the reward model (RM) in approximating the Llama-3 generator's utility (Eq.~\ref{eq:utility}) on the NewsQA validation set. To assess cross-architecture transfer, we benchmark the DeBERTa-based RM against general-purpose retrievers (BM25, BGE, and BGE-Reranker). As shown in Figure~\ref{fig:signal_analysis}, while standard models are frequently misled by semantic distractors (NDCG@1 $\le$ 0.72), RM achieves 0.86 NDCG@1 and 0.70 pairwise accuracy. This confirms RM as a high-fidelity proxy that successfully distills the heavy LLM's preferences into a compact supervisor. We set a utility threshold of 0.1 (calibrated to the top 10\% of the score distribution) for relevance. While LLM utility may diverge from human labels, a known trait in LLM-as-a-judge frameworks, we prioritize generator-specific alignment to ensure the retriever surfaces contexts that the fixed LLM can effectively utilize, which is the core objective of utility-aligned RAG.
\vspace{1mm}

\begin{figure}[t]
    \centering
    \vspace{-0.3cm}
    \includegraphics[width=\columnwidth]{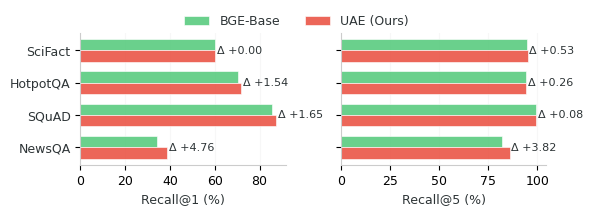}
    \vspace{-3mm}
    \caption{Zero-shot transfer performance. The model was trained \textit{only} on QASPER and evaluated on unseen datasets. UAE improves or retains performance across all domains.}
    \label{fig:transfer_results}
    \vspace{-5mm}
\end{figure}

\noindent{\bf Main Results.} Table \ref{tab:main_results} presents a comprehensive evaluation of UAE against three distinct categories of baselines: classic lexical retrievers (e.g., BM25, SPLADE), state-of-the-art dense retrievers (e.g., BGE, NV-Embed), and computation-heavy multi-stage rerankers (e.g., RankGPT, SePer). We analyze performance across three key dimensions:

\noindent\textit{Retrieval Performance.} UAE significantly outperforms all single-stage retrievers across both datasets. On the long-document QASPER benchmark, UAE achieves a MAP of 49.49, surpassing the strongest dense baseline (BGE-Base, 38.02) by $\approx11$ points, the late-interaction model ColBERTv2 (33.98) by $\approx15$ points and even outperforming the massive LLM-based embedder NV-Embed V2 (38.67). Notably, on NewsQA, UAE's retrieval quality (Recall@1 54.90) exceeds that of the computationally expensive RankGPT (49.68), demonstrating that aligning embeddings with generative utility can yield reranker-level precision in a single retrieval step. This advantage extends to ExpUtil@1 (average utility of the top-1 context). On NewsQA, UAE (5.818) surpasses both BGE-Base (4.738) and even the computation-heavy RankGPT (4.968), confirming that UAE prioritizes contexts maximally conducive to generation rather than mere semantic relevance.

\noindent{\textit{Generation Quality.}} These gains in retrieval translate directly to downstream generation fidelity. On NewsQA, UAE achieves a Gen-F1 of 45.6, surpassing standard dense retrieval (BGE-Base: 38.6) and outperforming the multi-stage RankGPT pipeline (43.7). Similarly, on QASPER, UAE improves Gen-F1 scores over classic baselines (27.0 vs. 23.9 for BGE), confirming that the retriever selects contexts that are not merely semantically relevant, but factually sufficient for the LLM to generate correct answers.

\noindent\textit{Efficiency \& Compatibility.} A critical barrier for deploying RAG in real-time production environments is latency; HCI research establishes that system response times must remain under 100 ms to be perceived as instantaneous by users \cite{responseMiller1968}. While multi-stage rerankers like RankGPT and SePer offer strong performance, they incur prohibitive latencies ($>$1600 ms), rendering them unsuitable for interactive applications. In contrast, UAE maintains a latency of $\approx$ 9 ms, matching the speed of standard bi-encoders. Furthermore, UAE is not mutually exclusive with reranking; as shown in the ablation study, combining UAE with rerankers (e.g., UAE + SePer) yields new state-of-the-art results (MAP 53.39 on QASPER), proving that UAE serves as a superior "first-stage" retriever that enhances the entire pipeline when latency budgets permit.\\

\noindent{\bf Zero-Shot Generalization and Robustness.} A common failure mode of domain-specific fine-tuning is catastrophic forgetting, where the model loses its ability to generalize to new tasks. To evaluate this, our QASPER-trained model is tested against four out-of-domain datasets: NewsQA , SQuAD \cite{rajpurkar-etal-2016-squad}, HotpotQA \cite{yang2018hotpotqa}, and SciFact \cite{wadden-etal-2020-fact}. Figure \ref{fig:transfer_results} summarizes the results. 

\section{Conclusion}

We present Utility-Aligned Embeddings (UAE), a framework that bridges the gap between the efficiency of dense retrieval and the high performance of utility-based models through distribution matching. Our results demonstrate that UAE significantly reduces semantic distractors and improves generation quality while operating 180$\times$ faster than LLM-based re-ranking methods. By maintaining standard ANN compatibility and serving as a high-quality foundation for multi-stage pipelines, UAE provides a practical and scalable solution for utility-driven RAG systems.

%%
%% The next two lines define the bibliography style to be used, and
%% the bibliography file.
\bibliographystyle{ACM-Reference-Format}
\bibliography{utility_bib}

\end{document}